# Accelerating Machine Learning Training Time for Limit Order Book Prediction

Mark J. Bennett
Department of Business Analytics
University of Iowa
Iowa City, Iowa 52242
mark-j-bennett@uiowa.edu

June 17, 2022

## ABSTRACT

Financial firms are interested in simulation to discover whether a given algorithm involving financial machine learning will operate profitably. While many versions of this type of algorithm have been published recently by researchers, the focus herein is on a particular machine learning training project due to the explainable nature and the availability of high frequency market data. For this task, hardware acceleration is expected to speed up the time required for the financial machine learning researcher to obtain the results. As the majority of the time can be spent in classifier training, there is interest in faster training steps. A published Limit Order Book algorithm for predicting stock market direction is our subject, and the machine learning training process can be time-intensive especially when considering the iterative nature of model development. To remedy this, we deploy Graphical Processing Units (GPUs) produced by NVIDIA available in the data center where the computer architecture is geared to parallel high-speed arithmetic operations. In the studied configuration, this leads to significantly faster training time allowing more efficient and extensive model development.

## 1 Introduction

High frequency trading (HFT) on a nanosecond time basis has been around for many years now and is beginning to dominate the volume in the equities and foreign exchange markets. Market participants and researchers have been concerned with future market direction, especially since 2018, as the machine learning approaches have been advancing. The Limit Order Book (LOB) of various volume depths on the bid and ask side of the market is the way for a matching engine to organize bids and offers for many traded securities. Of course, as buy-side firms attempt to buy and sell in the goal to maintain an overall positive profit, prediction becomes an advantage in the strategy. We discuss the advancements in LOB prediction first, then focus on studies of a specific computer program which is applied to NASDAQ security limit order books on widely traded securities such as Google (GOOG) as quoted with the LOBSTER limit order book price and volume dataset [1] which has been created for the academic economic research audience.

From experience with financial machine learning, the disadvantage of using prior mid-market prices exclusively in the non-stationary price time series is their random "jumpy" nature. Using merely a history of former prices can force machine learning classifiers to keep the direction and magnitude of the predictions of the future as different from today very close to zero. In other words, the classifier under-predicts with that limited data. Fortunately, with HFT, the LOB provides a deeper set of market features in order for the classifiers to use when minimizing their training and test loss function, and it is to their advantage to look outside of merely the recent mid-market price history.

The high frequency LOB-based price prediction problem is gaining momentum in the research community. Decision trees have proved to be a very popular and useful approach to machine learning in categorical and numeric prediction. Their advancement into bootstrapped aggregation with low inter-tree correlation is the basis of random forests (RF). Other decision approaches are

Gradient Boosted tree (GBMethod) and Extreme Gradient Boosted trees (XGBoost). As this research project evolved, there is a rich body of interesting results reported in [2,3,4,5,6,7,8,9,10,11].

This study is an adaptation of extensive practical LOB prediction work by Dr. Faisai Qureshi of the University of Ontario Institute of Technology, published in 2018 [11]. The associated Python application was found to be operational and intuitively implements in the aspect of the algorithms and the dataset. The dataset, if kept in original form, makes the code base and the market direction labeling easy to follow. So the goal of this work here is to report the improvement in terms of acceleration of the training process using a popular advanced computer architecture known as the Graphical Processing Unit (GPU) in order to reduce the iteration time during the effort of discovering the prediction solution.

The more mature Random Forest and other decision tree-based approaches are being eclipsed lately in the research community by the surge in the size of NLP classifiers such as neural networks and, specifically, attention-based transformers. This is understandable. Transformers are testing the limits of AI-based computing systems and extending the application-specific machine architectures. Developments such as that have focussed on news articles triggering market events and are still nascent with respect to the recent LOB prediction approaches. In spite of this trend, at times more mature classifiers can perform very well in certain data domains such as this LOB use case where purely numeric data is contained in the dataset.

**Market Modeling**

One of the most difficult challenges for market participants is predicting the market direction. With LOBs, this problem becomes more manageable because of the vectors of bid and ask prices and volumes accompanying the prices. In a level-10 LOB, we have prices and volumes for the bid and ask side of the market. For $L$ levels, a given point in time $t$:

$$b_{t,1}, ..., b_{t,L}, a_{t,1}, ..., a_{t,L}$$

are the bid and ask prices at those levels. When we are at time $t$ and we have a *mid-quote price* $p_t$ which is based upon the simple mean of the level-1 or highest bid $b_{t,1}$ and level-1 or lowest ask $a_{t,1}$ of any level-$l$ bid or ask volume size $V^b_{t,l}$ or $V^a_{t,l}$ and will fluctuate as time $t$ moves on. Mid-quoted prices are represented by the output or $p_{t+\Delta t}$ variable that we intend to predict the direction of in the next time period $t + \Delta t$.

$$p_t = (b_{t,1} + a_{t,1})/2$$

This mid-quote price above is a simpler version of what is known as the Volume-Weighted Average Price, VWAP, $w_t$ and also note that it is not volume weighted.

$$w_t = \frac{\sum_{l=1}^{L}\left(b_{t,l} \cdot V^b_{t,l} + a_{t,l} \cdot V^a_{t,l}\right)}{\sum_{l=1}^{L}\left(V^b_{t,l} + V^a_{t,l}\right)}$$

Up until now in this paper, the concern has been about the actual market mid-quote price data movement, but there are also well-known models where we simulate price data movement theoretically. Comparing these will help with discussing the direction prediction problem. In order to perform a prediction on the direction, there needs to be a consistent way to label the direction with discrete values. Actual market security price movement and simulated security price movement are both important to understand this labeling.

In financial analytics, price prediction has always presented a challenge and the price movement *direction* is one of the hardest pieces [12]. Qualitatively, one can think of thousands of market participants studying historical performance and news articles about securities and taking trading actions to maximize their economic profit. The resulting market price direction is unpredictable. This can be illustrated with a well-known market model. Quantitatively, the assumption of Geometric Brownian Motion for price behavior is well-known and forms the basis of the Black-Scholes-Merton approach to European option pricing. The model creators received the Nobel Prize in Economics in 1997, and it has been widely used as a method to generally determine the value of derivatives in dynamic markets [13]. If we follow this path, in longer term views, the markets have limited predictability and fall into the time series category of being considered non-stationary [12] due to the presence of random jump events. Looking to the model here, the focus is on the behavior of the underlying security.

Some models, such as those based upon GBM mentioned earlier, assume that distance from $p_t$ and $p_{t+\Delta t}$ for a time $t + \Delta t$ in the future, a figure known also as $\Delta p$ is proportional to the annualized historical volatility of the underlying security and also the square root of the distance from $t$ to $t + \Delta t$. This distance is $\Delta t$. For actual security prices, if our initial price is 100 dollars, the amount of price movement from this price $p_t$ to a new price $p_{t+\Delta t}$, the more time that evolves, the more the price can move away from price $p_t$.

For an intuitive model simulation example [14], using $\Delta t$ as 1 week of the 52 weeks in a year, $\Delta t = 1/52$. If the annualized volatility is 30% $=.30$ and our initial stock price $p_t$ is 100 dollars then the large change in price is $100 \cdot (.30 \cdot \sqrt{1/52} \cdot \epsilon)$ where $\epsilon$ is drawn from the normal distribution N(0,1). For $\epsilon = 3$ then $100 \cdot (.30 \cdot \sqrt{1/52} \cdot 3)$ we have a result of $\Delta p = 12.48$ dollars, a large upward movement in the price as our scenario and if $\epsilon =- 3$ then $100 \cdot (.30 \cdot \sqrt{1/52} \cdot (-3))$ and we have a result of $\Delta p =- 12.48$ dollars, a large downward movement. We can also surmise that if $\epsilon = 0$ then there is no movement. So, in summary, for the 3 cases where $\epsilon$ is in $\{3, 0, -3\}$ there are outcomes for $\Delta p$ which are $\{12.48, 0, -12.48\}$.

For the actual security prices from the LOB dataset here, in the corresponding actual upward price scenario, the direction is labeled +1 and in the downward scenario the direction is labeled -1, otherwise there is the no movement scenario when the change in price is zero and this direction is labeled 0. So the labels are selected from the set $\{+1, -1, 0\}$. According to the model, the reason that price prediction and price direction prediction is difficult is due to the presence of the $\epsilon$ random term. This term is modeled as any real value in the approximate range of [-3,+3] and to some extent outside of that range.

**Market Direction Labeling**

Sitting at time $t$ and moving ahead to a future time to $t + \Delta t$, successful trading involves a good prediction of the price movement from $p_t$ to $p_{t+\Delta t}$ more times than not. We define labels, down, up, and unchanged, as:

$$-1 \text{ when } p_{t+\Delta t} < p_t$$
$$+1 \text{ when } p_{t+\Delta t} > p_t$$
$$0 \text{ otherwise}$$

Python code was created to visualize the LOB and mid-quote price direction. Figure 1 has a two-frame plot of the LOB for the stock GOOG at two times since midnight marked to the 9-digit or nanosecond level behind the decimal point, with time abbreviated "tm" in the heading. In the middle of both plots is the mid-quote price marked as a vertical red line. We can see that the mid-quote prices jump quite a bit from the narrow market of the first plot to the wider market of the second plot.

In Figure 2, the label signal (blue) uses equal sized steps and is one step ahead of the actual market signal in order to train our machine learning classifiers as to the LOB conditions that formulate the next sample direction being $+1, -1, 0$. In general the orange line is a plot of the mid-quote prices and the blue line is an accumulation of the $+1, -1, 0$ so only take on integer values such as 577, 578, 579 dollars per share in this plot. In practice, we can ask the classifier to predict $\hat{y}_{t+\Delta t}$ the direction of the next next price $p_{t+\Delta t}$.

**Random Forest Classifiers**

We can think of a classifier for predicting LOB direction as a prediction function [15]. Random forests perform as one of the best types of classier in the study published in [11]. The goal is to reduce the variance of the values coming from the prediction function. With random forests, bootstrap aggregation, also known as bagging, can reduce this variance. Examining the relationship among the trees in the forest, the goal with a random forest is to minimize the correlation $\rho$ between the decision trees. If $B$ is the number of trees in the forest and $\sigma^2$ is the variance among $B$ correlated and identically distributed (i.d.) random variables with positive correlation $\rho$, then the variance of the average is

$$\rho\sigma^2 + \frac{1-\rho}{B}\sigma^2$$

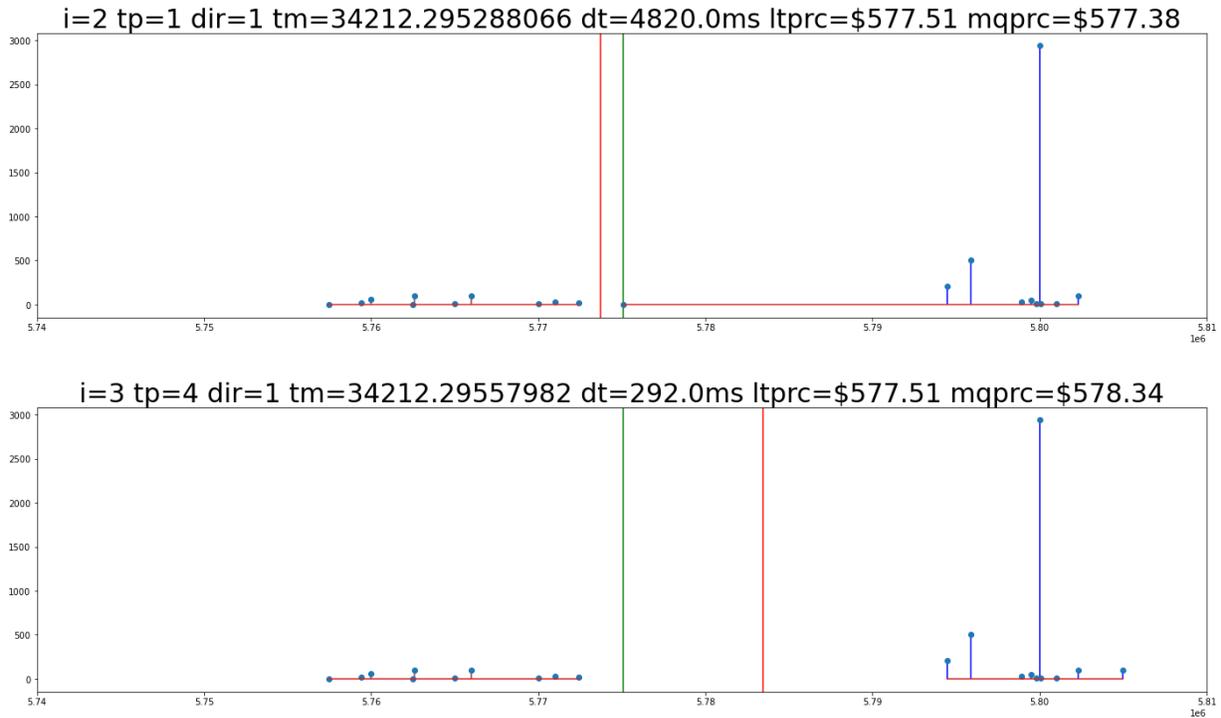

**Figure 1.** Limit order book snapshot for GOOG at two time points 292 milliseconds apart. Mid-quote price direction is upward almost a US Dollar in the mid-quote price as it goes from $577.38 to $578.34 as represented as the red line equally spaced between the rightmost of the connected bid points on the left and the leftmost of the connected ask points on the right in each plot. The x-axis represents the price levels and the y-axis represents the volume at each price.

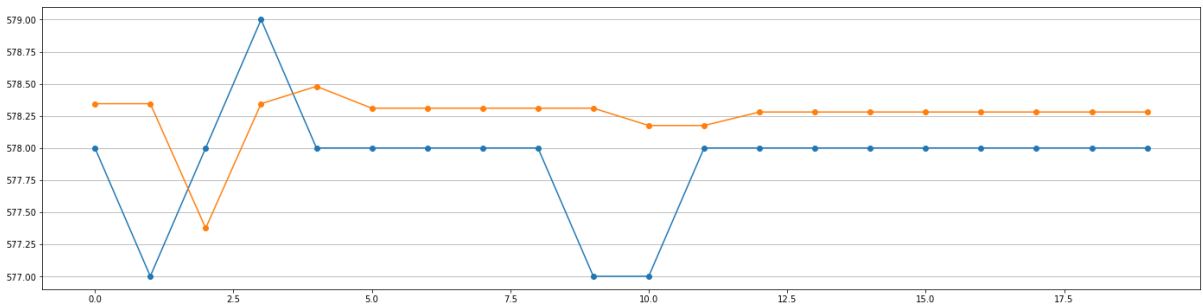

**Figure 2.** Mid-price quotes form a signal over time (orange). Labeling of the next time period market mid-price quotes with respect to the current period can be done using the set $\{+1, -1, 0\}$. When viewed in aggregate, these labels accumulate to form another signal over time (blue), where time spacing is forced to be equal to simplify the plot where the x-axis is the record index labeled "i" in Figure 1. The jump in price from $577.38 to $578.34 appears at x-values 2 and 3, for example.

Note that when *B* is large then the second term disappears and we are left with a small figure when ρ is close to zero. In summary, having a larger forest due to large *B*, then the variance of the average is smaller.

**GPU Improvement in Training Time**

Additional Python code was written to move the random forest from Sci-kit learn (SkLearn) into the Rapids CUML and CUPY hardware accelerated machine learning framework. Only a few lines of code were needed for this step. The enhanced Python code was run in the University of Iowa's High Performance Computing (HPC) environment on the Interactive Data Analytics Service (IDAS) research network. After multiple runs, the random forest model development time was discovered to be vastly improved with GPU acceleration. Without the GPU, the training time is 572.47 seconds whereas using an NVIDIA RTX 2080 Ti the training time is 26.23 seconds for a speedup of 22 X. The configuration has 4352 CUDA cores, CUDA Toolkit version 10.2, 11GB of GDDR6 memory, and CUDA driver version 440.44 using the CUML version of randomForest. With most runs being 22 times faster, iteration time improves for model training which was the subject of our investigation: whether massive parallel processing would speed up the machine learning training steps

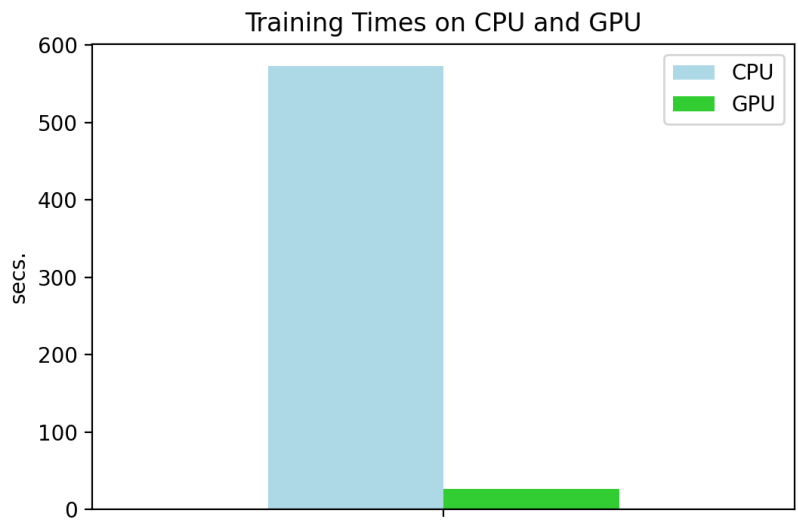

**Figure 3.** Training times for the CPU and then the GPU: lower is better.

22 X speedup per training run can certainly accumulate during the model development cycle. Furthermore, from experience, model development training time would be expected to go down to 50% of the time on an RTX 2080 Ti by using a larger data center GPU, either a Volta model V100 or an Ampere model A100, due in part to the number of CUDA cores doubling over the RTX 2080 Ti. The Python-based Rapids package was used for converting data frames into GPU-based data frames with CUML providing the GPU machine learning classifiers [16].

**Multiple GPU Configuration**

Training times reported here are for a single security. As many market practitioners know, the "beat" of every market security is slightly different. GOOG trades differently than AMZN in terms of the price jumps and volatility. By using multiple GPUs, multiple securities can be handled by multiple corresponding classifiers also running on multiple GPUs. By assigning a security and its LOB market dataset, machine learning training runs can get assigned to a GPU in a pool, and then multiple GPUs can be deployed, providing faster training, with each GPU receiving consecutive training requests, and fulfilling them per security and dataset. This is the approach in [17] for 115 Nasdaq stocks for the period January 1, 2019 through January 31, 2020 where an entire supercomputing environment of 332 NVIDIA GPUs exists to expand machine learning training across the important stocks.

**Conclusions**

The goal of this study is to examine the possibility of hardware acceleration of LOB prediction for high frequency trading. After many investigations of approaches from the reference list, the approach documented in [11] served as the basis of our acceleration experiment. Python routines and the LOBSTER dataset are quite useful for academic economic research studies of this nature and it includes high profile securities such as Google (GOOG) and other highly liquid large technology stocks. After defining the up and down market direction labeling, machine learning techniques can be applied to the 10-level bid and ask dataset for mid-quote price prediction. For the training runs, when moving the data frames from Pandas to Rapids, the timing of the CPU-only and GPU-based runs were reported for comparison to see the amount of engineering time improvement.

———————————————